**The revised SNIP indicator of Elsevier's Scopus**

Dear Sir,

The modified SNIP indicator of Elsevier, as recently explained by Waltman *et al*. (2013) in this journal, solves some of the problems which Leydesdorff & Opthof (2010 and 2011) indicated in relation to the original SNIP indicator (Moed, 2010 and 2011). Our major objection was that the SNIP value was obtained by dividing an arithmetic average in the numerator by a median value in the denominator which results in a quotient without the possibility of specifying the statistical uncertainty in the indicator. Instead of dividing aggregates, one should normalize observed against expected values in terms of distributions. This discussion followed our discussion of the "crown indicator" (Opthof & Leydesdorff, 2010) and led to an increasing awareness in the community that one should not use "rates of averages" (RoA), but instead average the rates (AoR) when normalizing values against the expected ones so that valid statistical error measures can be applied (Gingras & Larivière, 2011).

The recent modifications to SNIP elegantly solve the problem of normalization by using the same premises as the original indicator, but in the appendix the authors are able to prove that the revised SNIP indicator can also be considered as an arithmetic average (Equation A2, at p. 281). However, the authors also mention that the use of arithmetic averages as expected values is unfortunate in the case of scientometric distributions because these can be extremely skewed (Seglen, 1992 and 1997). Somewhat amazingly therefore, the authors did not consider the possibility of using the median (instead of the mean) or other non-parametric statistics as these same authors have acknowledged in other publications (e.g., Waltman *et al.*, 2012; cf. Leydesdorff & Bornmann, 2011 and 2012).

Except for providing Elsevier with an improved alternative to the impact factor (IF) as a measure used by its main competitor (Thomson Reuters), the added value to the evaluation process of this new indicator appears limited because it is virtually impossible to control as an outsider the value of the revised SNIP indicator—or equally Elsevier's revised Science Journal Ranking (SJR2; cf. Guerrero-Bote & Moya-Anegón, 2012)—even if one has online access to the Scopus database. One main objection to the use of the IF has been its unreliability in terms of data collection; but the data could easily be checked independently using online facilities. In some cases, the producer revised the indicator in response to such criticism (e.g., Brumback, 2008; McVeigh & Mann, 2009; cf. Moed & Van Leeuwen, 1996).

Since journal indicators are often used in evaluation processes of individual scholars and groups, the transparency to and possible reproduction by the evaluated scholars may be relevant for current debates about the ethics of scientometrics (e.g., at http://citationculture.wordpress.com/2013/07/29/bibliometrics-of-individual-researchers/; cf. Spaan, 2010). The new generation of journal indicators is increasingly opaque to external observation except in terms of formal procedures. Leaving outliers (e.g., *Acta Crystallographica Section A*) apart, the revised SNIP of *JAMA*, for example, is more than 40% higher than the original one (after proper normalization using Eq. 6, at p. 280),[1] whereas the SNIP of the *International Journal of Computer Vision* is 5.74 against 14.64 for the original SNIP (this is a difference of 5.90 or more than 100% of the revised value). However, the authors provide some reasons why this latter decrease can perhaps be expected.

---

[1] The revised SNIP value of *JAMA* is provided in Table 3 as 9.37, whereas a difference of 3.83 (= 40.1%) is listed for this journal in Table 5. The original and revised values of SNIP for the *International Journal of Computer Vision* are provided in Tables 4 and 3, respectively.

A trend towards formalization and therefore black-boxing of the data may be unavoidable with further professionalization, but the publication of this revised Elsevier indicator by a leading group of evaluators emblematically affiliated with Elsevier (for example, at <http://www.socialsciences.leiden.edu/cwts/news/new-scopus-journal-classification.html>)[2] makes us also aware of the thin lines between bibliometrics and commercial objectives. Such thin lines may have become structural for innovative systems (Gibbons et al., 1994; Leydesdorff & Zawdie, 2010), but may also set limitations to the possibilities to change a new version of a previously introduced indicator. If possible, in my opinion, one should use a non-parametric statistics for the bibliometric evaluation (e.g., Leydesdorff, 2012; Wagner & Leydesdorff, 2012) and use indicators that are sufficiently transparent to be reproduced independently.

Loet Leydesdorff,
Amsterdam School of Communication Research, University of Amsterdam,
Kloveniersburgwal 48, 1012 CX Amsterdam, The Netherlands;
loet@leydesdorff.net

---

[2] At this website, one formulates: "Elsevier and CWTS have a long-standing tradition of collaboration. Over 20 years, CWTS has developed and discussed tool and methods for Elsevier to meet user needs. A recent information product CWTS developed for Scopus is the impact measure for journals (SNIP)."